\documentclass[aps,prd,superscriptaddress,showpacs,preprint,nofootinbib]{revtex4}
\usepackage{graphicx, bm}
\usepackage{amsmath}
\usepackage{mathrsfs}

\usepackage[usenames]{color}
\usepackage{float}

\begin{document}
\draft

\title{Stellar energy loss rates beyond the standard model}

\author{A. Llamas-Bugar\'in\footnote{maria.llamas@fisica.uaz.edu.mx}}
\affiliation{\small Unidad Acad\'{e}mica de F\'{\i}sica, Universidad Aut\'onoma de Zacatecas\\
	Apartado Postal C-580, 98060 Zacatecas, M\'exico.\\}

\author{A. Guti\'errez-Rodr\'{\i}guez\footnote{alexgu@fisica.uaz.edu.mx}}
\affiliation{\small Unidad Acad\'{e}mica de F\'{\i}sica, Universidad Aut\'onoma de Zacatecas\\
	Apartado Postal C-580, 98060 Zacatecas, M\'exico.\\}

\affiliation{\small Unidad Acad\'emica de Estudios Nucleares, Universidad Aut\'onoma de Zacatecas,
         98060 Zacatecas, M\'exico.\\}

\author{ A. Gonz\'alez-S\'anchez}
\affiliation{\small Unidad Acad\'{e}mica de Ciencia y Tecnolog\'{\i}a de la Luz y la Materia,
              Km. 6, C.P. 98160, Zacatecas, M\'exico.\\}

\author{ M. A. Hern\'andez-Ru\'{\i}z\footnote{mahernan@uaz.edu.mx}}
\affiliation{\small Unidad Acad\'emica de Ciencias Qu\'{\i}micas, Universidad Aut\'onoma de Zacatecas\\
         Apartado Postal C-585, 98060 Zacatecas, M\'exico.\\}

\author{ A. Espinoza-Garrido}
\affiliation{\small Unidad Acad\'{e}mica de Ciencia y Tecnolog\'{\i}a de la Luz y la Materia,
              Km. 6, C.P. 98160, Zacatecas, M\'exico.\\}

\author{ A. Chubikalo}
\affiliation{\small Unidad Acad\'{e}mica de Ciencia y Tecnolog\'{\i}a de la Luz y la Materia,
              Km. 6, C.P. 98160, Zacatecas, M\'exico.\\}

\date{\today}

\begin{abstract}

It is known that the dipole moments of the neutrino lead to important astrophysical and cosmological effects.  In this regard,
within the context of a $U(1)_{B-L}$ model, we develop and present novel analytical formulas to assess the effects of the anomalous
magnetic moment and electric dipole moment of the neutrino on the stellar energy loss rates through some common physical process of
pair-annihilation $e^+e^-\to(\gamma, Z, Z^{\prime})\to\nu\bar\nu$. Our results show that the stellar energy loss rates strongly depends
on the effective magnetic moment of the neutrino, but also on the parameters which characterize the adopted $U(1)_{B-L}$ model.
\end{abstract}

\pacs{14.60.St, 13.40.Em, 12.15.Mm \\
Keywords: Non-standard-model neutrinos, Electric and Magnetic Moments, Neutral Currents.}

\vspace{5mm}

\maketitle

\section{Introduction}

At the beginning of the last century, the establishment of the Hertzsprungh-Russell (HR) diagram led quickly to the bases of a stellar evolution theory, which identified mass as the fundamental parameter to determine the structure, nucleosynthesis, mean lifetime, ways of production and release of energy, stability mechanisms and the final product of stars. We now know that massive stars with minimum initial mass of approximately $7 M_{\odot} < M < 9 M_{\odot}$ are the potential precursors of different types of supernovae type I (SN I) with subtypes Ia, Ib, and Ic. The former two being mainly linked to single white dwarfs, moderate mass interacting binaries, whereas the lattest of them seems to be linked to massive Wolf-Rayet stars \cite{Kilpatrick2018}. Supernovae type II (SN II), with subtypes IIn, IIP, IIL IIb and IIe are associated with massive stars of initial masses within $10 M_{\odot} < M < 35 M_{\odot}$, resulted from red supergiant stars, most massive white dwarfs and very luminous blue variable stars \cite{hendry2006,smartt2009}. Each of these SN populations are potential progenitors of neutron stars and black holes.

In a SN event, the violent outburst triggered by the gravitational collapse of a Fe-core of mass approximately $1.3-2M_{\odot}$ toward a neutron star, releases an energy about $E_{\nu}\approx (3-5)\times 10^{53}$ erg in the form of neutrinos and antineutrinos of the three flavors ($\nu_e, \nu_\mu, \nu_\tau$). The gravitational binding energy of the resulting neutron star corresponds nearly to ($10-15\%$) of the core rest mass. The remaining core-collapsed SN can reach extremely high surface temperatures and densities, {\it e.g.} up to $3\times 10^8 K$ and $\rho\le10^5 gr/cm^3$. Therefore, contrary to normal stars where photons can be transported to the surface by radiative or convective processes, neutrinos interact extremely weakly and can easily escape from stellar interiors taking away such amounts of energy. These stellar energy loss rates, which leads to cooling by neutrino emission, are mainly due to purely leptonic processes and weak nuclear reactions. However although large values of temperatures and densities seems to characterize well the final stage of stellar evolution, the former are dominant. Under these conditions, we can identify four main mechanisms of neutrino pair production relevant for the neutrino luminosity \cite{beaudet1967,dicus1972,dicus1976,alam1989,dicus1977,bruenn1985,torres2010,hernandez2017}:

\begin{align}
e^{+}e^{-}&\rightarrow \nu \bar \nu  \hspace{3mm}  \textrm{    (pair-annihilation)}, \label{pair}\\
\gamma + e^{\pm} &\rightarrow e^{\pm}\nu \bar \nu  \hspace{3mm}  \textrm{    ($\nu -$ pair photoproduction)},\\
\gamma^* &\rightarrow \nu \bar \nu  \hspace{3mm}   \textrm{    (plasmon decay)},\\
e^{\pm} +Z &\rightarrow  e^{\pm} +Z \nu \bar \nu  \hspace{3mm}   \textrm{    (bremsstrahlung)}.
\end{align}

Quantify stellar energy loss rates is recognized as a priority scientific knowledge to set constraints on the properties and interaction of light particles
\cite{dicus1972,dicus1976,ellis1983,bruenn1985}, which give important advances in
astrophysics and cosmology \cite{tanabashi2018,gerbino2017,raffelt1996}. Furthermore, one of the most interesting possibilities to use stars as particle physics laboratories is to study the backreaction of the novel energy loss rates implied by the existence of new low-mass particles such as axions \cite{payez2015,fischer2016}, which remain as candidates to dark matter.

On the other hand, the weak interaction predicts interactions between the neutrino and photon through a non-zero magnetic moment
induced via loop corrections of gauge boson. In the minimal extension of the Standard Model (SM) with massive Dirac neutrinos,
the diagonal magnetic moment of the neutrino mass eigenestate $\nu_i$ is proportional to its mass $m_i$ and
is given by \cite{Fujikawa-MM,Robert-MM,Fukugita1-MM,bugarin2017}:

\begin{equation}
\mu^{ii}_\nu=\frac{3eG_F m_i}{8\sqrt{2}\pi^2}= 3.2\times 10^{-19}(\frac{m_i}{eV})\mu_B,
\end{equation}

\noindent where $\mu_B$ is the Bohr magneton.

Discoveries of new interactions and/or nonstandard properties of neutrinos could modify the rate at which neutrinos are produced,
or they can even carry more, or less energy. In this respect, for instance, the electromagnetic properties of the neutrino, such
as the Anomalous Magnetic Moment (AMM) and the Electric Dipole Moment (EDM) have given one of the most sensitive probes of physics
beyond the Standard Model (BSM) \cite{mohapatra2004,bugarin2017,ko2019}. These interactions are expected to generate observable
effects in astrophysical environments \cite{kerimov1992,blinnikov1994,raffelt1999,heger2009,giunti2015}. Furthermore, they will
be of relevance in cosmology \cite{gerbino2017}, and terrestrial neutrino experiments \cite{cisneros1971}.

Bounds on the diagonal magnetic moment of the neutrino have been reported as the result of different astrophysical and experimental determinations; for example, \cite{Barbieri-MM} show that the abscence of high-energy events in the SN1987A neutrino signal puts a higher bound of $\mu_\nu \lesssim 10^{-12}\mu_B$ at $90\%$. Supernova energy loss studies lead to an upper limit of
$\mu_\nu \lesssim (1.1-2.7)\times 10^{-12}\mu_B$ at $90\%$ according to \cite{Kuznetsov-MM}. Cooling rates of red giants
\cite{Raffelt-MM} gives a comparable bound of $\mu_\nu \lesssim 3\times 10^{-12}\mu_B$ at $90\%$, whereas analysis of cooling
rates of white dwarfs \cite{Blinnikov-MM} puts a limit of $\mu_\nu \lesssim 10^{-11}\mu_B$ at $90\%$. Recent bounds were also obtained
for the Borexino experiment \cite{Borexino-MM} which explores solar neutrinos $\mu_{\nu} \leq 5.4\times 10^{-11}\mu_B$ at $90\%$, and the TEXONO Collaboration \cite{Texono-MM} get $\mu_{\nu} < 2.9\times 10^{-11}\mu_B$ at $90\%$, etc. All these bounds are reasonably compatible.

Concerning the simplest extensions to the SM, the $U(1)_{B-L}$ model \cite{mohapatra1980,marshak1980, buchmuller1991,emam2007,khalil2008}
take into account an extra $U(1)$ local gauge symmetry \cite{carlson1987}, where B and L denote the baryon number and lepton number,
respectively. The B-L model \cite{basso2009} is attractive due to its relatively simple theoretical structure, as it contains an extra
gauge boson $Z^{\prime}$ corresponding to B-L gauge symmetry and an extra SM singlet scalar (heavy Higgs boson $H$). The mass of the
additional boson $Z^{\prime}$ is given by the relation $M_{Z^{\prime}}=2v^{\prime}g^{\prime}_1$ \cite{khalil2008,emam2007,basso2009}.
This boson $Z^{\prime}$ interacts with the leptons, quarks, heavy neutrinos and light neutrinos with interaction strengths proportional
to the B-L gauge coupling $g^{\prime}_1$. The $U(1)_{B-L}$ model becomes mostly attractive due that can account for dark matter abundance \cite{singirala2018}.

It is important to point out that, in the early Universe, from a few femtoseconds after the Big-Bang up to the electroweak scale,
period in which the last positrons disappeared, the relevant degrees of freedom are pressure, the energy density, entropy density,
and number density. Then, in the B-L model neutrinos are not produced in purely left-handed states, probably the thermally excited
degrees of freedom in the early universe would increase, providing constraints on the alternative model.
A review of the fundamentals of the minimal B-L model adopted in this paper can be read in \cite{bugarin2017} and references therein.

The priority objective of this paper is to study the combined effects of {\it a)} the AMM and the EDM of the neutrino, and {\it b)} the parameters of the $U(1)_{B-L}$ model on the determination of stellar energy loss rates through the process of pair-annihilation $e^+e^- \to (\gamma, Z, Z^{\prime}) \to \nu_i \bar\nu_i$ with $\nu_i=\nu_e, \nu_\mu, \nu_\tau$. We shall present analytical exact formules for this purpose, and we get the relative emission compared with those of the SM.

With these goals, the paper is organized as follows: Within the context of the $U(1)_{B-L}$ model, in Section II we perform the calculations of the transition amplitude, the total cross-section and the exact determination of the stellar energy loss rates. The relative correction for the stellar
energy loss rates with the SM is also shown. In Section III, we present a discussion and our results. Finally, we give our main conclusions in Section IV.

\section{Stellar energy loss rates through the process $e^{+}e^{-} \to \nu \bar \nu$}

It is already well known that during the later phases of stellar evolution, once the core bounce and a proton-neutron star forms at the center of the SN explosion, the huge amount of energy of the order of $10^{53}$ ergs is realised in the form of pairs neutrinos and antineutrinos. Evidently, neutrinos of the three flavors with extra freedom degrees such as AMM and EDM would be able to take away more amounts of energy. This is the case of the neutrino involved in the pair-annihilation process $e^{+}e^{-}\rightarrow\nu\bar\nu$ given by Eq. \eqref{pair} which would be an important agent for cooling the collapsed core by neutrino emission.

In our calculations, we incorporate the AMM and the EDM of the neutrino. We also quantify the dependence on these momenta of the stellar energy loss rate through the process of pair-annihilation $e^+e^- \to (\gamma, Z, Z^{\prime}) \to \nu \bar\nu$ and on the parameters associated with the $U(1)_{B-L}$ model.

\subsection{Amplitudes calculation}

\begin{figure}[H]
\centerline{\scalebox{.7}{\includegraphics{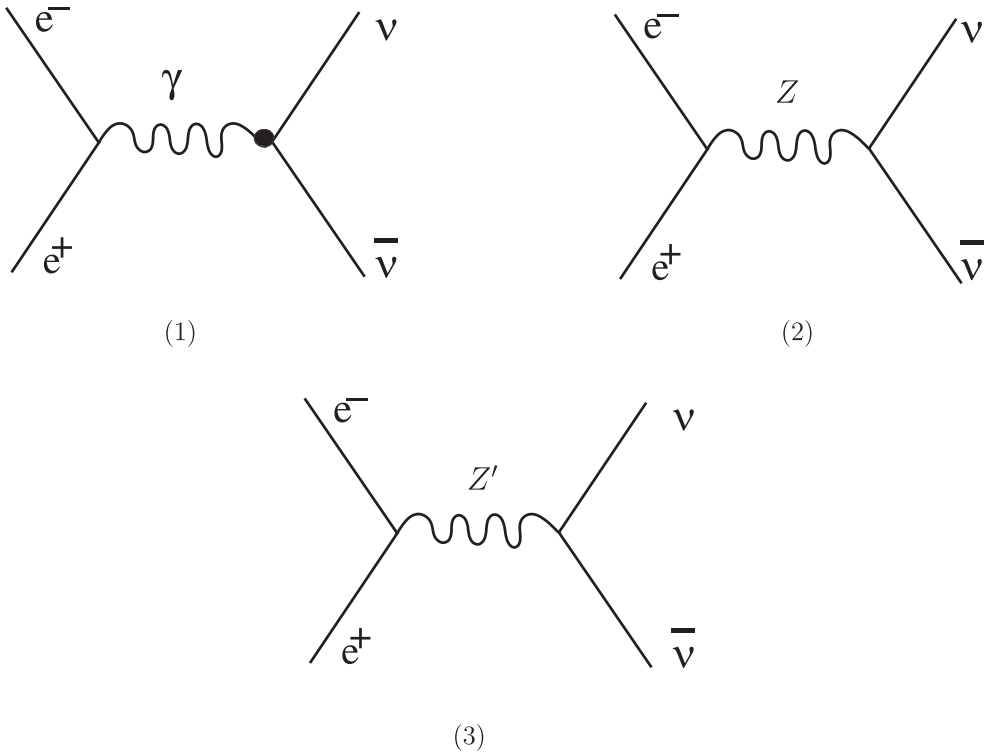}}}
\caption{ \label{fig:FeyDiagrams} The Feynman diagrams contributing to the process
$e^{+}e^{-}\rightarrow (\gamma,Z,Z^{\prime}) \to \nu \bar \nu$. The dot represents
an interaction arising from a effective operator.}
\end{figure}

We focus now on the calculation of the amplitude for the pair-annihilation:

\begin{equation}
e^{+}\left(p_1 \right) e^{-}\left(p_2 \right)
\rightarrow (\gamma, Z, Z^{\prime})\to \nu_i\left(k_1,\lambda_1 \right)\bar\nu_i\left(k_2,\lambda_2 \right),
\label{process}
\end{equation}

\noindent with $\gamma$, $Z$, $Z^{\prime}$ exchanges and $\nu_i=\nu_e, \nu_\mu, \nu_\tau$, respectively. Here the $ k_i $
and $ p_i $ are the particles quadrimoment and $\lambda_i$ is the neutrino helicity. The Feynman diagrams contributing
to the process given by Eq. (\ref{process}) are shown in Fig. \ref{fig:FeyDiagrams}.

For neutrinos the electric charge is zero and there are no electromagnetic interactions at tree-level between the
neutrino field $\nu(x)$ and the electromagnetic field $A^\mu(x)$. However, such interactions can arise at the quantum
level from loop diagrams at higher order of the perturbative expansion of the interaction. In the one-photon approximation,
the electromagnetic interactions of a neutrino field $\nu(x)$ can be described by an effective interaction. In this regard,
the electromagnetic properties of any fermion appear in quantum field theory through its interaction with the photon. Thus,
the most general expression for the effective vertex of the interaction $ \nu\bar{\nu}\gamma$ is given by \cite{nieves1982,kayser1983}:

\begin{equation}
\Gamma^{\alpha}=eF_{1}(q^{2})\gamma^{\alpha}+\frac{ie}{2m_{\nu}}F_{2}(q^{2})\sigma^{\alpha\mu}q_{\mu}+\dfrac{e}{2m_\nu}F_3(q^2)\gamma_5\sigma^{\alpha\mu}q_\mu +eF_4(q^2)\gamma_5(\gamma^\alpha -\dfrac{q\llap{/}q^\alpha}{q^{2}}),
\end{equation}

\noindent where $q^\mu$ is the photon momentum, and $F_{1, 2, 3, 4}(q^2)$ are the electromagnetic form factors of the neutrino. Strictly, the $F_{1, 2, 3, 4}(q^2)$ form factors are not physical quantities, but in the limit $q^2\to 0$ they become quantifiable and related to the static quantities corresponding to Charge Radius, AMM, EDM and the Anapole Moment (AM), respectively \cite{giunti2015,vogel1989}. In this analysis, we are interested in the AMM and the EDM of the neutrino, which are defined in terms of the $ F_2\left(q^2=0\right) $ and $ F_3\left( q^2=0\right) $ form factors as follows \cite{gutierrez2018}:

\begin{equation}
\mu_\nu=\frac{m_e}{m_\nu}F_2\left(q^2=0\right),  \qquad d_\nu=\dfrac{e}{2m_\nu} F_3\left(q^2=0\right).
\end{equation}

\noindent Throughout the rest of the paper, the form factors corresponding to charge radius and the anapole moment are not considered.

Using the couplings between the $ Z $, $Z^{\prime}$ bosons with those of the SM fermions given in Table \ref{tabla:couplings} of Appendix A, the respective transition amplitudes are thus given by:

\begin{align}
i\mathcal{M}_\gamma&=\bar{u}\left(k_2,\lambda_2\right)\Gamma^{\alpha}\upsilon\left(k_1,\lambda_1\right) \dfrac{-ig_{\mu\nu}}{\left(p_1+p_2 \right)^2}
\bar{\upsilon}\left(p_1\right)ie\gamma_\mu u\left(p_2\right),\\
i\mathcal{M}_Z&= \bar{u}\left(k_2,\lambda_2\right)\left( \dfrac{-ig}{\cos{\theta_W}}\right)\gamma^\mu \frac{1}{2}\left(g_V^\nu-{g_A^\nu\gamma}_5\right)\upsilon\left(k_1,\lambda_1\right)\nonumber \\
&\quad\times \dfrac{-i(g_{\mu\nu}-p_\mu p_\nu/M_{Z}^2)}{\left[\left(p_1+p_2\right)^2 -M_Z^2-iM_Z\Gamma_Z\right]}  \bar{\upsilon}\left(p_1\right)\left( \dfrac{-ig}{\cos{\theta_W}}\right)\gamma^\nu\frac{1}{2}\left(g_V^e-{g_A^e\gamma}_5\right)u\left(p_2\right),\\
i\mathcal{M}_{Z^{\prime}}&=\bar{u}\left(k_2,\lambda_2\right)\left( \dfrac{-ig}{\cos{\theta_W}}\right)\gamma^\mu \frac{1}{2}\left(g^{\prime\nu}_V-{g^{\prime\nu}_A\gamma}_5\right)\upsilon\left(k_1,\lambda_1\right)\nonumber \\
&\quad\times \dfrac{-i(g_{\mu\nu}-p_\mu p_\nu/M_{Z^\prime}^2)}{\left[\left(p_1+p_2\right)^2 -M_{Z^\prime}^2-iM_{Z^\prime}\Gamma_{Z^\prime}\right]} \bar{\upsilon}\left(p_1\right)\left( \dfrac{-ig}{\cos{\theta_W}}\right)\gamma^\nu\dfrac{1}{2}\left(g^{\prime e}_V-{g^{\prime e}_A\gamma}_5\right)u \left(p_2\right),
\end{align}

\noindent where $ u $ and $ v $ are the usual Dirac spinors, and the electron and positron helicity indexes have been suppressed since they will be averaged over. The constants $g_V^e$, $g_A^e$, $g_V^\nu$ and $g_A^\nu$ depend only of the parameters of the $U(1)_{B-L}$ model, that is $ \theta_{B-L} $ and $g^{\prime}_1$ (see Table \ref{tabla:couplings} of Appendix A).

The squared transition amplitude is derived by the use of Eqs. (9)-(11), resulting:

\begin{align}
\sum_{s}&\left|\mathcal{M}_{\nu\bar\nu}\right|^2=\dfrac{4\left(4\pi \alpha\right)^2 }{\sin^4 {2\theta_W}}\Bigg\lbrace \dfrac{\sin^4 {2\theta_W} }{\left(4\pi \alpha\right)}\left(\mu_\nu^2 + d_\nu^2 \right)\left(\dfrac{\left(p_1\cdot p_2 +m_e^{2}\right)\left(p_1\cdot k_2+m_e^{2} \right) }{\left(p_1+p_2 \right)^2 } \right
)\nonumber \\
&+\Bigg[\bigg(\dfrac{1}{M_Z^2}\big(g_V^e-g_A^e \big)\big( g_V^\nu +g_A^\nu\big)+\dfrac{1}{M_{Z^\prime}^2}\big(g_V^{\prime e}-g_A^{\prime e} \big)\big( g_V^{\prime \nu} +g_A^{\prime\nu}\big)\bigg)^2\nonumber \\
&+\bigg(\dfrac{1}{M_Z^2}\big(g_V^e+g_A^e \big)\big( g_V^\nu -g_A^\nu\big)+\dfrac{1}{M_{Z^\prime}^2}\big(g_V^{\prime e}+g_A^{\prime e} \big)\big( g_V^{\prime \nu} -g_A^{\prime\nu}\big)\bigg)^2\Bigg]\left(p_1\cdot k_1 \right)\left(p_2\cdot k_2 \right) \nonumber \\
&+\Bigg[\bigg(\dfrac{1}{M_Z^2}\big(g_V^e+g_A^e \big)\big( g_V^\nu +g_A^\nu\big)+\dfrac{1}{M_{Z^\prime}^2}\big(g_V^{\prime e}+g_A^{\prime e} \big)\big( g_V^{\prime \nu} +g_A^{\prime\nu}\big)\bigg)^2\nonumber \\
&+\bigg(\dfrac{1}{M_Z^2}\big(g_V^e-g_A^e \big)\big( g_V^\nu -g_A^\nu\big)+\dfrac{1}{M_{Z^\prime}^2}\big(g_V^{\prime e}-g_A^{\prime e} \big)\big( g_V^{\prime \nu} -g_A^{\prime\nu}\big)\bigg)^2\Bigg]\left(p_1\cdot k_2 \right)\left(p_2\cdot k_1 \right) \nonumber \\
&+2\Bigg[\dfrac{1}{M_Z^4}\Big( \left(g_V^{e}\right)^2 -\left(g_A^{e}\right)^2 \Big)\Big( \left(g_V^{\nu}\right)^2 +\left(g_A^{\nu}\right)^2 \Big)+\dfrac{1}{M_{Z^\prime}^4}\Big( \left(g_V^{\prime e}\right)^2 -\left(g_A^{\prime e}\right)^2 \Big)\nonumber \\
&\times\Big( \left(g_V^{\prime\nu}\right)^2 +\left(g_A^{\prime\nu}\right)^2 \Big) +\dfrac{2}{M_Z^2M_{Z^\prime}^2}\big(g_V^{e}g_V^{\prime e}-g_A^{e}g_A^{\prime e}\big)\big( g_V^{\nu}g_V^{\prime \nu}+g_A^{\nu}g_A^{\prime \nu}\big)\Bigg]\left(m_e^{2}\right)\left(k_1\cdot k_2 \right)\Bigg\rbrace,
\label{Amplitude}
\end{align}

\noindent where $ s=\left(p_1+p_2 \right) ^2 $, $ t=\left(p_1-k_1 \right) ^2 $, $ u=\left(p_2-k_2 \right) ^2 $ and $ s+t+u=2m_e^2 $ are the Mandelstam variables. The coupling constants of Eq. \eqref{Amplitude} are redefined as presented in Appendix B.

The SM expression for the squared transition amplitude of the same process can be recovered
in the decoupling limit when $ \theta_{B-L}=0 $, $g^{\prime}_1=0$, $ M_{Z^{\prime}}\to
\infty $ and $ \mu_{\nu}=d_{\nu}=0 $. In this case, the terms that depend on
$ \theta_{B-L} $, $g^{\prime}_1$, $ M_{Z^{\prime}}$, $\mu_{\nu}$ and $d_{\nu} $
in Eq. \eqref{Amplitude} are zero and Eq. \eqref{Amplitude}  is reduced to the SM case
\cite{dicus1972,ellis1983,yakovlev2001,esposito2002,esposito2003}.

\subsection{Stellar energy loss rates}

The expression for the stellar energy loss rates for our pair-annihilation process $e^+e^- \to (\gamma, Z, Z') \to \nu \bar\nu$
is determined by \cite{kerimov1992,yakovlev2001,esposito2002,heger2009}:

\begin{equation}
\mathcal{Q}_{\nu\bar{\nu}}^{B-L}=\frac{4}{\left(2\pi\right)^6}\int_{m_e}^{\infty}{\frac{d^3\mathbf{p}_1}{\left[ e^{\left(E_1-\mu\right)/T}+1\right] }\frac{d^3\mathbf{p}_2}{\left[ e^{\left(E_2+\mu\right)/T}+1\right] }\left(E_1E_2\right)}\upsilon_{rel}
\sigma_{Tot}^{B-L},
\label{loss}
\end{equation}

\noindent as a function of the total spin-averaged cross-section of the process $\sigma_{Tot}^{B-L}$, the Fermi-Dirac distribution
functions  $\left[ \exp\left( {\left(E_{1,2}\pm\mu\right)/T}\right)+1\right]^{-1} $ for $ e^{\pm} $, the chemical potential,
$\mu$, for the electron, the stellar temperature $T$, and $\upsilon_{rel}$ is the electron-positron relative velocity $\frac{1}{2}[s(s-4m_e^{2})]^{1/2}$ \cite{aydin1992}.

The quantity $ E_1E_2\upsilon_{rel} \sigma_{Tot}^{B-L}$ is Lorentz invariant and is given by \cite{heger2009}

\begin{eqnarray}
E_1E_2\upsilon_{rel}
\sigma_{Tot}^{B-L}&=&\dfrac{1}{4}\int{\dfrac{d^3k_1d^3k_2}{\left( 2\pi\right)^{3}2\omega_1 \left( 2\pi\right)^{3}2\omega_2 }} \left|\mathcal{M}_{\nu\bar\nu}\right|^2\left( 2\pi\right)^{4}\delta^4\left(p_1+p_2-k_1-k_2 \right).
\end{eqnarray}

\noindent For process \eqref{process}

\begin{align}
{E_1E_2\upsilon_{rel}}\sigma_{Tot}^{B-L}&=\dfrac{\pi \alpha^2 }{3\sin^4 {2\theta_W}}\Bigg\lbrace\dfrac{\sin^4\theta_{W}}{2\pi\alpha}\left(\mu_\nu^2 + d_\nu^2 \right)\left(2m_e^2+p_1\cdot p_2\right)+\Big(g_{1}^{[B-L]}\Big)\Big[m_e^4\nonumber \\
&+3m_e^2\left(p_1\cdot p_2 \right)+2\left(p_1\cdot p_2 \right)^2  \Big]+12\Big(g_{2}^{[B-L]}\Big)\Big[m_e^4+m_e^2\left(p_1\cdot p_2 \right)\Big]\Bigg\rbrace,
\end{align}

\noindent where the coefficients $g_{1,2}^{[B-L]}$ contain the couplings of the $ U(1)_{B-L} $ model and are given in Appendix B.

The calculation of the stellar energy loss rates given by Eq. \eqref{loss} can be easily performed by expressing the latest integrals
in terms of the Fermi integral, defined as

\begin{equation}
G_n^\pm(\lambda,\eta,x)=\lambda^{{3+2n}}\int_{\lambda^{-1}}^{\infty}{x^{2n+1}\frac{\sqrt{x^2-\lambda^{-2}}}{e^{\left(x\pm\eta\right)}+1}}dx, \label{Fermi}
\end{equation}

\noindent where we have defined the dimensionless variables

\begin{equation}
\lambda=\dfrac{k_{B} T}{m_e},  \qquad \eta=\dfrac{\mu}{k_BT},\nonumber
\end{equation}

\noindent and we take $ k_{B}=1 $ for the Boltzmann constant. With these definitions, Eq. (16) becomes

\begin{equation}
G_s^\pm(\lambda,\eta,E)=\dfrac{1}{m_e^{3+2s}}\int_{m_e/T}^{\infty}{E^{2s+1}\frac{\sqrt{E^2-m_e^{2}}}{e^{\left(E\pm\mu_e\right)/T}+1}}dE.
\end{equation}

Therefore, the final expression for the stellar energy loss rates, in the context of the $U\left(1 \right)_{B-L} $ model is given by:

\begin{align}
&\mathcal{Q}_{\nu\bar{\nu}}^{B-L}{\tiny \left(\mu_{\nu}, d_{\nu},\theta_{B-L}, g^{\prime}_{1},M_{Z^{\prime}},\eta  \right)}=\dfrac{\alpha^2 m_e^9}{9\pi^3\sin^4 {2\theta_W}}\Bigg\lbrace\dfrac{3\sin^4{2\theta_{W}}}{2\pi\alpha m_e^2}\left(\mu_\nu^2 + d_\nu^2 \right)\nonumber \\
&\qquad \qquad\qquad \times\bigg[2\Big(G_{-1/2}^-G_{0}^+ + G_{0}^-G_{-1/2}^+\Big)+G_{0}^-G_{1/2}^+ +G_{1/2}^-G_{0}^+\bigg]\nonumber \\
&\qquad \qquad+4\Big(g_{1}^{[B-L]}\Big) \bigg[5\Big(G_{-1/2}^-G_{0}^+ + G_{0}^-G_{-1/2}^+\Big) +7\Big(G_{0}^-G_{1/2}^+ +G_{1/2}^-G_{0}^+\Big)\nonumber \\
&\qquad \qquad\qquad -2\Big(G_{1}^-G_{-1/2}^+ +G_{-1/2}^-G_{1}^+\Big)+8\Big(G_{1}^-G_{1/2}^+ +G_{1/2}^-G_{1}^+\Big) \bigg]\nonumber \\
&\qquad\qquad +36\Big(g_{2}^{[B-L]}\Big)\bigg[G_{-1/2}^-G_{0}^+ + G_{0}^-G_{-1/2}^+ +G_{0}^-G_{1/2}^+ +G_{1/2}^-G_{0}^+\bigg]\Bigg\rbrace.
\label{emissivity}
\end{align}

Note, that the weak and electromagnetic diagrams given in Fig. 1 do not interfere. This equation is valid whether or not the electrons
are degenerate or even relativistic, {\it i.e} is valid for all values of the $\lambda $ and $\eta $. Additionally, one can once again
recover the SM stellar energy loss rates in the decoupling limit \cite{dicus1972}.

We emphasize that, while the dependence of mixing angle $ \theta_{B-L} $ between $Z-Z^{\prime} $ and the coupling constant $ g_1^{\prime} $ of the $ U(1)_{B-L}$ model are contained in the new coupling constants $ g^{f}_V $, $ g^{f}_A $, $ g^{\prime f}_V $ and $ g^{\prime f}_A $ (see Appendix A), also the dependence on the $ \eta $ degeneration parameter is contained in the Fermi integrals $ G_n^{\pm}\left(\lambda,\eta \right)  $. Thus, to quantify the combined effects of the AMM and the EMM of the nuetrino, with the free parameters $\theta_{B-L}$, $g'_1$ and $M_{Z'}$, of the $ U\left(1 \right)_{B-L} $ model, we define the relative correction for the stellar energy loss rates as:

\begin{equation}
\dfrac{\delta \mathcal{Q}_{\nu\bar\nu}^{B-L} }{\mathcal{Q}_{\nu\bar\nu}^{SM}}=\dfrac{\mathcal{Q}_{\nu\bar\nu}^{B-L}\left(\mu_{\nu}, d_{\nu},\theta_{B-L}, g^{\prime} _1,M_{Z^{\prime}},\eta  \right) - \mathcal{Q}_{\nu\bar\nu}^{SM}\left(\eta \right)  }{\mathcal{Q}_{\nu\bar\nu}^{SM}\left(\eta \right) }.
\label{Correction}
\end{equation}

However, given the mathematical impossibility of solving these integral for all the values of $\lambda$ and $\eta$ in general,
we proceed to evaluate Eq. \eqref{emissivity} in different interesting limits represented by their respective densities
and temperatures. We shall do this next.

\subsubsection{Region I: $\lambda\ll 1, \eta\ll\frac{1}{\lambda} $}

This corresponds to the nonrelativistic and nondegenerate case, where temperature and
densities can vary between $3\times10^8 \le T \le 3\times10^9K$ and $\rho\le10^5 gr/cm^3$,
respectively. In this regime, the Fermi integral is simplified as:

\begin{equation}
G_n^\pm \approx\left(\frac{\pi}{2}\right)^{\frac{1}{2}}\lambda^{\frac{3}{2}}e^{-1/\lambda}e^{\mp\eta},
\end{equation}

\noindent which allows us to get:

\begin{align}
\mathcal{Q}_{I}^{B-L}&\left(\mu_{\nu}, d_{\nu},\theta_{B-L},g^{\prime}_{1},M_{Z^{\prime}} \right)=\dfrac{2\alpha^2m_e^6}{\pi^2 \sin^4{2\theta_{W}}}\left(T\right)^{3} e^{-2m_{e}/T}\nonumber \\
&\times\Bigg[\dfrac{\sin^4{2\theta_{W}}}{4\pi\alpha m_e^2}\left(\mu_\nu^2 + d_\nu^2 \right)+4\Big(g_{1}^{[B-L]}+g_{2}^{[B-L]}\Big)\Bigg].
\label{R1}
\end{align}

Therefore, the relative correction gives:

\begin{eqnarray}
\dfrac{\delta \mathcal{Q}^{B-L}_{I} }{\mathcal{Q}^{SM}_{I}}&=&\dfrac{M_Z^{4}\bigg[ \Big(g_{1}^{[B-L]}+g_{2}^{[B-L]}\Big)+\dfrac{\sin^4{2\theta_{W}}\left(\mu_\nu^2 + d_\nu^2 \right)}{16 \pi\alpha m_e^2}\bigg] }{\Big( g_V^{SM}\Big)^2 } -1.
\label{C_RI}
\end{eqnarray}

The result is independent of the stellar temperature, and Eq. \eqref{C_RI} only depends on the AMM $ \left( \mu_{\nu}\right) $ and the EDM
$ \left( d_{\nu}\right) $ of the neutrino, and on the parameters of the $U(1)_{B-L}$ model, explicitly the coupling constant $ g^{\prime}_1 $
and the gauge boson mass $ M_{Z^{\prime}} $.

\subsubsection{Region II: $\lambda\ll1, \frac{1}{\lambda}\ll\eta\ll\frac{2}{\lambda} $}

Such a nonrelativistic and mildly degenerate case, represents temperatures $T\le10^8K$ and
densities between $10^4 gr/cm^3\le \rho \le 10^6 gr/cm^3$. Then, Fermi integrals satisfy
that $G_0^-\gg G_0^+$ and $G_n^- \approx G_0^-$, thus

\begin{equation}
G_n^+\approx G_0^+=\left(\frac{\pi}{2}\right)^{\frac{1}{2}}\lambda^{\frac{3}{2}}e^{-1/\lambda}e^{-\eta}, \qquad G_n^-\approx G_0^-=\left(\frac{\rho}{\mu_e}\right)\frac{\pi^2}{m_e^3}N_A.\\
\end{equation}

\noindent With this, we assess:

\begin{align}
\mathcal{Q}_{II}^{B-L}&\left(\mu_{\nu}, d_{\nu},\theta_{B-L},g^{\prime}_{1},M_{Z^{\prime}} \right)=\dfrac{2\sqrt{2\pi}\alpha^2} {\pi\sin^4{2\theta_{W}}}\left(\dfrac{\rho}{\mu_e} N_{A}\right)\left( \dfrac{T}{m_e}\right)^{3/2}{m_e^6}\,e^{-({m_{e}+\mu_e})/{T}} \nonumber\\
&\qquad\qquad\qquad\qquad \times\Bigg[\dfrac{\sin^4{2\theta_{W}}}{4\pi\alpha m_e^2}\left(\mu_\nu^2 + d_\nu^2 \right)+4\Big(g_{1}^{[B-L]}+g_{2}^{[B-L]}\Big)\Bigg],
\label{R2_1}
\end{align}

\noindent and the corresponding relative correction results:

\begin{eqnarray}
\dfrac{\delta \mathcal{Q}^{B-L}_{II} }{\mathcal{Q}^{SM}_{II}}&=&\dfrac{M_Z^{4}\bigg[ \Big(g_{1}^{[B-L]}+g_{2}^{[B-L]}\Big)+\dfrac{\sin^4{2\theta_{W}}\left(\mu_\nu^2 + d_\nu^2 \right)}{16 \pi\alpha m_e^2}\bigg] }{\Big( g_V^{SM}\Big)^2 } -1,
\label{C_RII}
\end{eqnarray}

\noindent which is exactly equal as in region I, extending this equality in the dependence on the bunch of parameters just listed above. Consequently, the indistinguishability of treating with non-degenerate or mildly degenerate electrons becomes clear.

\subsubsection{Region III: $\lambda\ll1, 1\ll\lambda\eta $}

This region represents the relativistic and degenerate case and is valid for temperatures
$T>6\times10^7K$ and densities $\rho>10^7 gr/cm^3$. Fermi integrals result:

\begin{equation}
G_n^+\approx G_0^+=\left(\frac{\pi}{2}\right)^{\frac{1}{2}}\lambda^{\frac{3}{2}}e^{-1/\lambda}e^{-\eta}, \qquad G_n^-=\left(\frac{3}{2n+3}\right)\left(\lambda\eta\right)^{2n}G_0^-.
\end{equation}

Then, the energy loss rates for this region to highest power in $ \lambda\eta $ is:

\begin{align}
\mathcal{Q}_{III}^{B-L}\left(\theta_{B-L},g^{\prime}_{1},M_{Z^{\prime}} \right)&=\dfrac{8\sqrt{2\pi}\alpha^2} {5\pi\sin^4{2\theta_{W}}}\left(\dfrac{\rho}{\mu_e} N_{A}\right)\left( \dfrac{T}{m_e}\right)^{3/2}\left(\dfrac{\mu_e}{m_e}\right)^2\nonumber\\
&\qquad\times{m_e^6}\,e^{-({m_{e}+\mu_e})/{T}} \Big(g_{1}^{[B-L]}\Big),
\label{R3_1}
\end{align}

\noindent and consequently the relative correction is given by:

\begin{eqnarray}
\dfrac{\delta \mathcal{Q}^{B-L}_{III} }{\mathcal{Q}^{SM}_{III}}&=&\dfrac{2(M_Z^4)\Big(g_{1}^{[B-L]}\Big)}{ \Big[\Big({g_V^{SM}}\Big)^{2}
+\Big(g_A^{SM}\Big)^{2} \Big] }-1.
\label{C_RIII}
\end{eqnarray}

The typical approximation for this region only considers the terms of dominant powers, so there is no dependence on the AMM and/or EDM of the neutrino.

\subsubsection{Region IV: $\lambda\gg1, \eta\ll1 $}

The relativistic and nondegenerate case holds for densities $\rho>10^7 gr/cm^3$. In this
region we may ignore the chemical potential. Considering the dominance of the highest orders in $\lambda$

\begin{equation}
G_n^\pm\approx \lambda^{2n+3} \Gamma\left(2n+3\right)\sum_{S=1}^{\infty}\frac{\left(-1\right)^{S+1}}{S^{2n+3}}.
\end{equation}

\noindent Then, the stellar energy loss rates for this region is:

\begin{align}
\mathcal{Q}_{IV}^{B-L}\left(\theta_{B-L},g^{\prime}_{1},M_{Z^{\prime}},\mu_\nu,d_\nu \right)&=\dfrac{28\pi \zeta{\left( 5\right) }\, \alpha^2\left( T\right)^{9}} {3\sin^4{2\theta_{W}}} \Big(g_{1}^{[B-L]}\Big),
\label{R4_1}
\end{align}

\noindent thus the relative correction is the same as in the previous region:

\begin{eqnarray}
\dfrac{\delta \mathcal{Q}^{B-L}_{IV} }{\mathcal{Q}^{SM}_{IV}}&=&\dfrac{(M_Z^4)\Big(g_{1}^{[B-L]}\Big)}{\Big[\left(g_V^{SM}\right)^{2} +\left(g_A^{SM}\right)^{2} \Big]}-1.
\label{C_RIV}
\end{eqnarray}

\subsubsection{Region V: $\lambda\gg1, \eta\gg1 $}

This degenerate relativistic region holds for densities greater than $\rho>10^8 gr/cm^3$
with temperatures of $ T\approx 10^{10}K$ at the lowest density, extendable to a range
between $10^{10}K$ and $10^{11}K$ at a density of $\rho>10^{10} gr/cm^3$. Here $G_n^-\gg G_n^+$, then

\begin{equation}
G_n^+\approx \lambda^{2n+3} \left(2n+2\right)! e^{-\eta} \qquad G_n^-\approx \left(\frac{3}{2n+3}\right)\left(\lambda\eta\right)^{2n}\left(\frac{\rho}{\mu_e}\right)\frac{\pi^2}{m_e^3}N_A.
\end{equation}

Restricting the calculation to the higher powers in $\lambda\eta$, the stellar energy loss
rates result:

\begin{align}
\mathcal{Q}^{B-L}_{V}\left(\theta_{B-L},g^{\prime}_{1},M_{Z^{\prime}},\rho\right)&=\dfrac{\left( 8\alpha\right) ^2}{5\pi \sin^4{2\theta_{W}}}\left(\dfrac{\rho}{\mu_e} N_{A}\right)\left( \dfrac{T}{m_e}\right)^{4}\left( \dfrac{\mu_e}{m_e}\right)^{2}\nonumber\\
&\qquad\times{m_e^6}\,e^{-\mu_e/{T}}\Big(g_{1}^{[B-L]}\Big),
\label{RV}
\end{align}

\noindent thus the relative correction is:

\begin{eqnarray}
\dfrac{\delta \mathcal{Q}^{B-L}_{V} }{\mathcal{Q}^{SM}_{V}}&=&\dfrac{2(M_Z^4)\Big(g_{1}^{[B-L]}\Big)}{\Big[ \Big({g_V^{SM}}\Big)^{2}
+\Big(g_A^{SM}\Big)^{2} \Big] }-1,
\label{C_RV}
\end{eqnarray}

\noindent resulting equal as in Region III. Again, it becomes clear that there is an indistinguishability of treating with nondegenerate
or degenerate electrons.

\section{Results}

In this paper in the context of the $SU(3)_C\times SU(2)_L\times U(1)_Y\times U(1)_{B-L}$ model \cite{mohapatra1980,marshak1980, buchmuller1991,emam2007,khalil2008,carlson1987,basso2009} we develop and present
novel analytical formulas to assess the effects of the anomalous magnetic moment and the electric dipole moment of the
neutrino, in addition of the parameters of the B-L model on the stellar energy loss rates through the physical process
of pair-annihilation $e^+ e^-\to(\gamma, Z, Z^{\prime})\to\nu\bar\nu$. This is one of the main mechanisms of neutrino
pair production relevant for the neutrino luminosity.

For nonvanishing AMM and/or EDM of neutrinos, the pair production process via the channel of Eq. \eqref{process} would receive
an additional electromagnetic contribution from the interaction with a virtual photon. Consequently, new channels for neutrino
production are possible and the additional electromagnetic pair production would lead to an increase in the stellar neutrinos.
In here, our analytical result on the stellar energy loss rates of neutrino pair production are obtained through the process
$e^+e^-\to(\gamma,Z,Z^{\prime})\to \nu \bar\nu$ and expressed as a function of the $\mu_{\nu}$ and the $ d_{\nu}  $ of the
neutrino, the mixing angle $\theta_{B-L} $, the mass of a new gauge boson $M_{Z^{\prime}}$, which at the time, strongly depends
on $ g_{1}^{\prime} $ and the degeneration parameter $ \eta $.

\begin{figure}[H]
	\centerline{\scalebox{.9}{\includegraphics{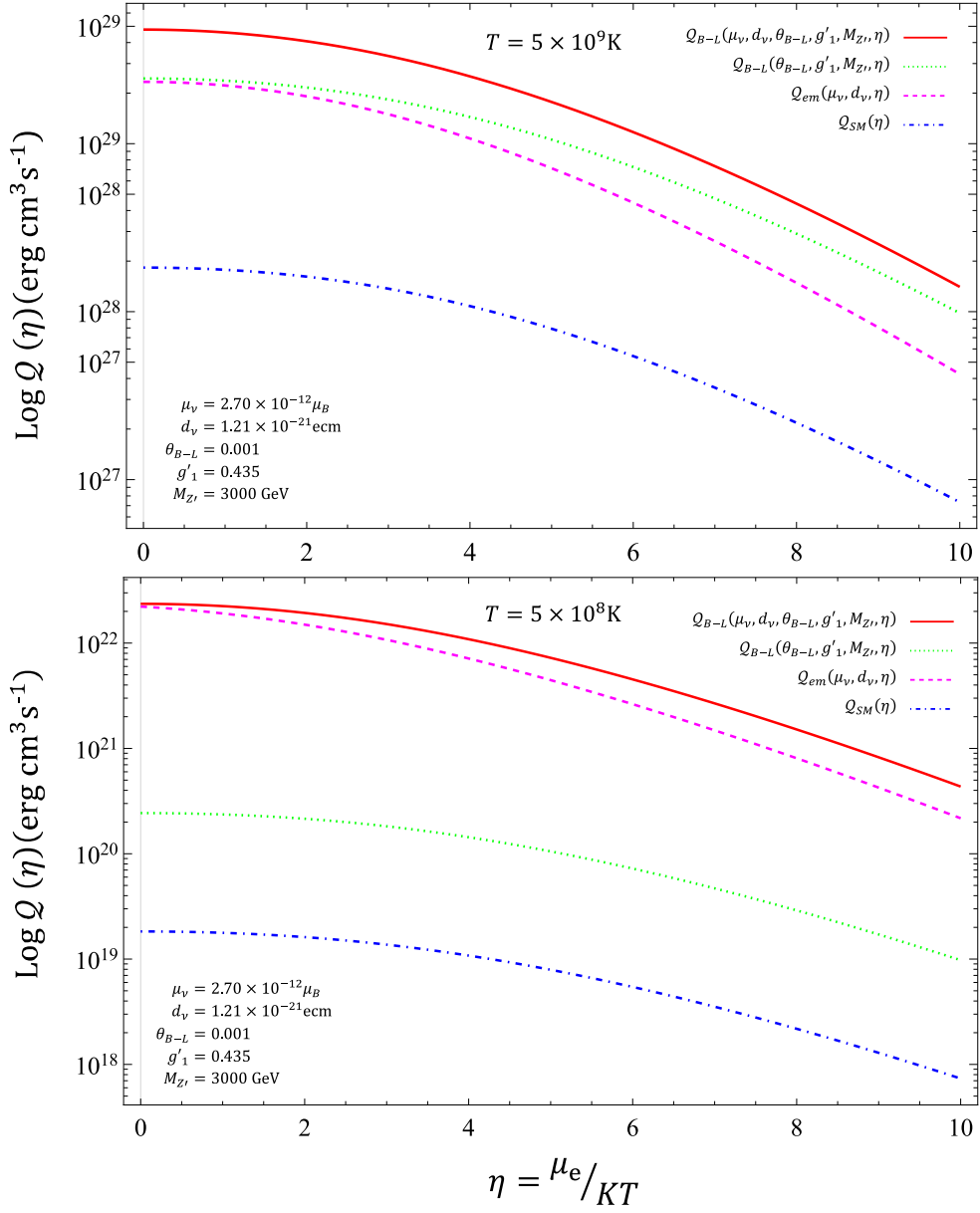}}}
	\caption{ \label{fig:Models} The stellar energy loss rates as a
		function of the degeneration parameter $\eta$ for stellar temperature of $ 5\times10^8$K and $ 5\times10^9 $K.
        The solid line is for $\mathcal{Q}_{B-L}\left(\mu_{\nu}, d_{\nu},\theta_{B-L},g^{\prime}_{1},M_{Z^{\prime}}, \eta \right)$,
		the tiny dashes line is for $\mathcal{Q}_{B-L}\left(\theta_{B-L},g^{\prime}_{1},M_{Z^{\prime}}, \eta \right)$, the large
        dashes line is for $\mathcal{Q}_{em}\left(\mu_{\nu}, d_{\nu}, \eta \right)$ and the dot-dashes line is for
        $\mathcal{Q}_{SM}\left(\eta \right)$.}
\end{figure}

The stellar energy loss rates for the process of pair-annihilation $e^+e^- \to (\gamma, Z, Z^{\prime}) \to \nu \bar\nu$ as a
function of the degeneration parameter $\eta $ are presented in Fig. \ref{fig:Models}, corresponding to input parameters
$g^{\prime}_1=0.435$, $M_{Z^{\prime}}=3000$ GeV, $\theta_{B-L}=10^{-3}$, $\mu_{\nu}=2.70\times 10^{-12}\mu_B$ \cite{Kuznetsov-MM}
and $d_{\nu}=1.21\times 10^{-21} ecm$ \cite{bugarin2017}. In this figure we consider the following scenarios, $\mathcal{Q}_{B-L}
\left(\mu_{\nu}, d_{\nu},\theta_{B-L},g^{\prime}_{1},M_{Z^{\prime}}, \eta \right)$ the stellar energy loss rates as a function
of the dipole moments of the neutrino, the parameters of the B-L model and the degeneration parameter $\eta$, that is to say the
$\mathcal{Q}_{B-L}$ correspond to the B-L model with electromagnetic properties. Another case corresponds to $\mathcal{Q}_{B-L}\left(\theta_{B-L},g^{\prime}_{1},M_{Z^{\prime}}, \eta \right)$ the stellar energy loss rates as a
function of the parameters of the B-L model and the degeneration parameter $\eta$, that is the $\mathcal{Q}_{B-L}$ correspond to the
$U(1)_{B-L}$ model and without electromagnetic properties of the neutrino. The third case correspond to $\mathcal{Q}_{em}\left(\mu_{\nu},
d_{\nu}, \eta \right)$ the stellar energy loss rates as a function of the dipole moments of the neutrino and the degeneration parameter
$\eta$, this case refers to the minimal extension of the SM with electromagnetic properties of the neutrino. Finally, the $\mathcal{Q}_{SM}
\left(\eta \right)$ corresponds to the case of SM. These results show that both the dipole moments and the parameters of the B-L model have
an important effect on the stellar energy loss rates. For instance, the difference between the cases with and without electromagnetic properties
for the B-L model is 2 orders of magnitude. For the cases of the SM and the minimally extended SM with electromagnetic properties the difference
is up to 3 orders of magnitude. Fig. \ref{fig:Models} reveals that emissivity becomes larger with the contribution of the AMM and of the EDM
of the neutrino, as well as with the parameters of the B-L model.

\begin{figure}[H]
	\centerline{\scalebox{.8}{\includegraphics{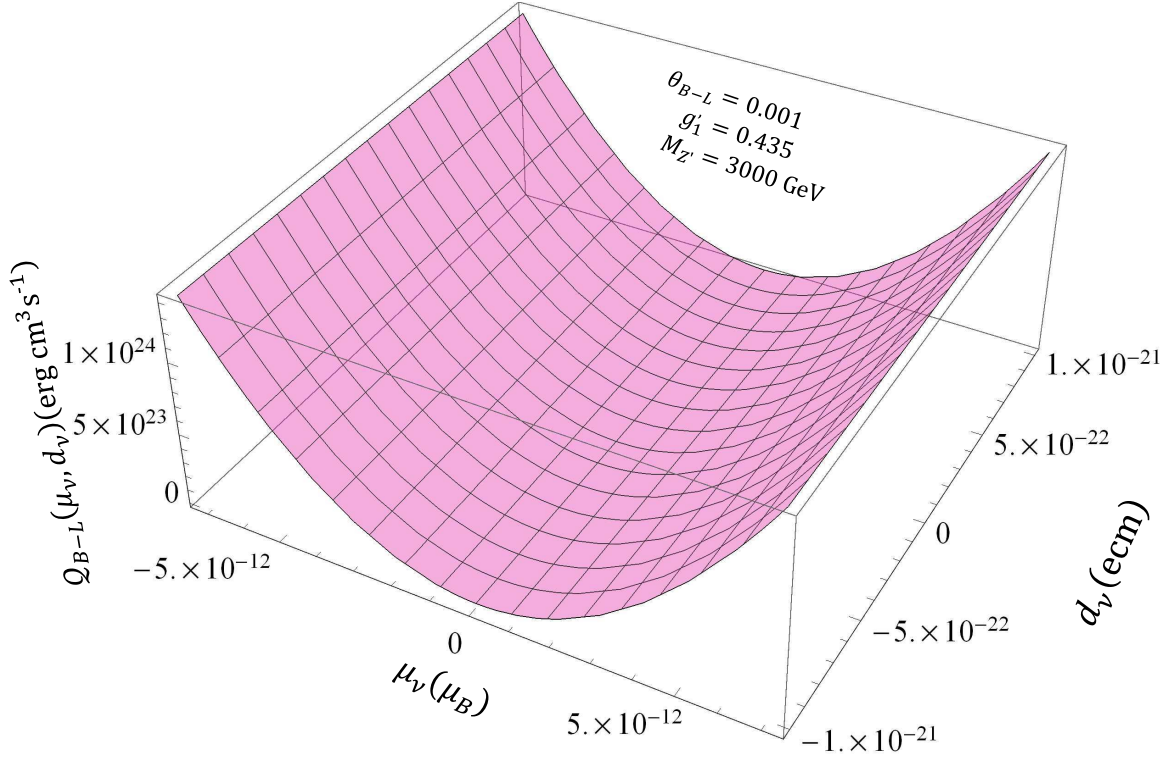}}}
	\caption{ \label{fig:MD} The stellar energy loss rates
		as a function of the AMM and the EDM $(\mu_{\nu}, d_{\nu})$ for $\eta =2$.}
\end{figure}

Variation of ${\cal Q}_{B-L}\left(\mu_\nu, d_{\nu}, \eta \right)$ as a function of $\mu_\nu$ and $d_{\nu}$ with $g^{\prime}_1=0.435$,
$M_{Z^{\prime}}=3000$ GeV, $\theta_{B-L}=10^{-3}$ and degeneration parameter $\eta=2$, respectively, is shown in Fig. \ref{fig:MD}.
This, clearly shows a strong dependence of ${\cal Q}_{B-L}\left(\mu_\nu, d_{\nu}, \eta \right)$ with respect to the AMM of the neutrino
and it is almost independent of the EDM, in agreement with a previous analysis presented in Ref. \cite{kerimov1992}.

\begin{figure}[H]
	\centerline{\scalebox{.9}{\includegraphics{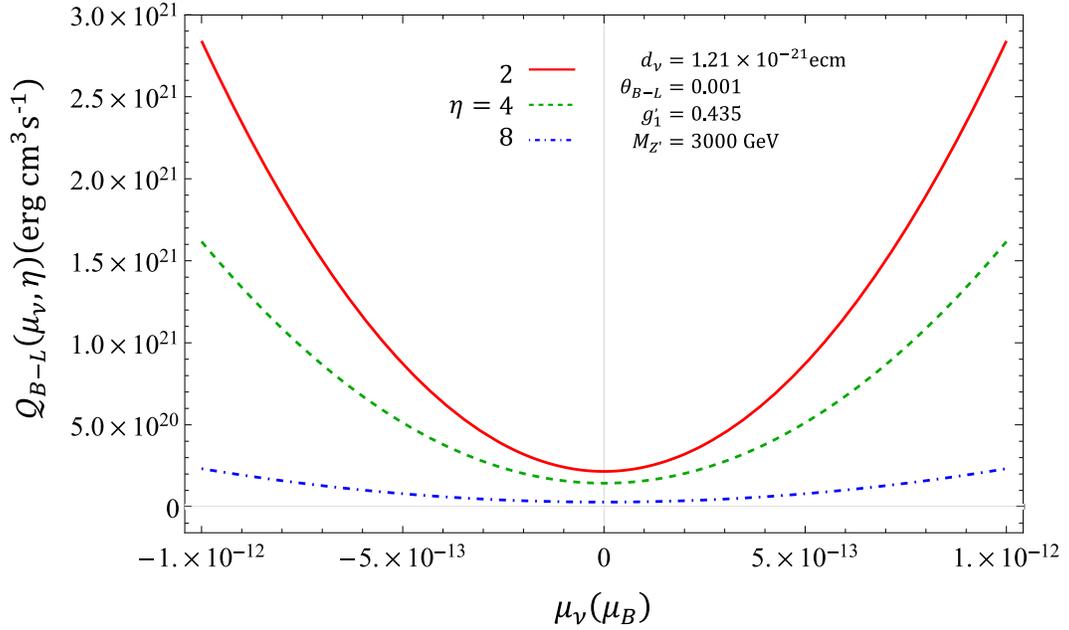}}}
	\caption{ \label{fig:MDM} The stellar energy loss rate
		as a function of the AMM of the neutrino. The solid line is for $\eta=2$, the dashes line is for $\eta=4$
and the dot-dashes line
		is for $\eta=8$.}
\end{figure}

The dependence of ${\cal Q}\left(\mu_{\nu}, \eta \right)$ with respect to $\mu_{\nu}$ is display in Fig. \ref{fig:MDM}, with
degeneration parameters $\eta= 2, 4,8$. From this figure we see that the stellar energy loss rates depend significantly on both,
the AMM and the degeneration parameter $\eta$. The stellar energy loss rates also decreases as increasing $\eta$, which is due
to the reduction in the number of positrons involved in the collisions.

\begin{figure}[H]
	\centerline{\scalebox{.7}{\includegraphics{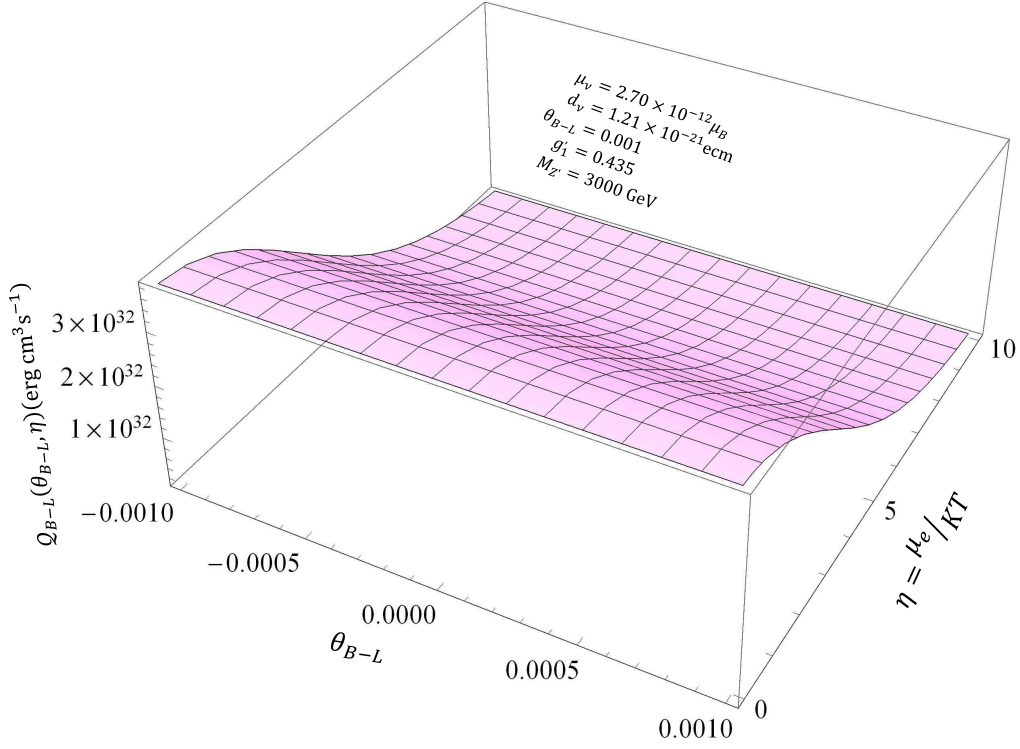}}}
	\caption{ \label{fig:B-L} The stellar energy loss rates
		as a function of degeneration parameter and the mixing angle $(\eta, \theta_{B-L}) $.}
\end{figure}

The dependence of ${\cal Q}\left(\theta_{B-L}, \eta \right)$ on the pair of parameters $(\theta_{B-L}, \eta)$ is shown in
Fig. \ref{fig:B-L}. The stellar energy loss rates decreases when $\eta $ increases. This behaviour is due to the reduction of
the number of positrons available to cause the collision. Besides, the stellar energy loss rates keeps nearly constant for any
value of the mixing angle $\theta_{B-L} $.

\begin{figure}[H]
	\centerline{\scalebox{.8}{\includegraphics{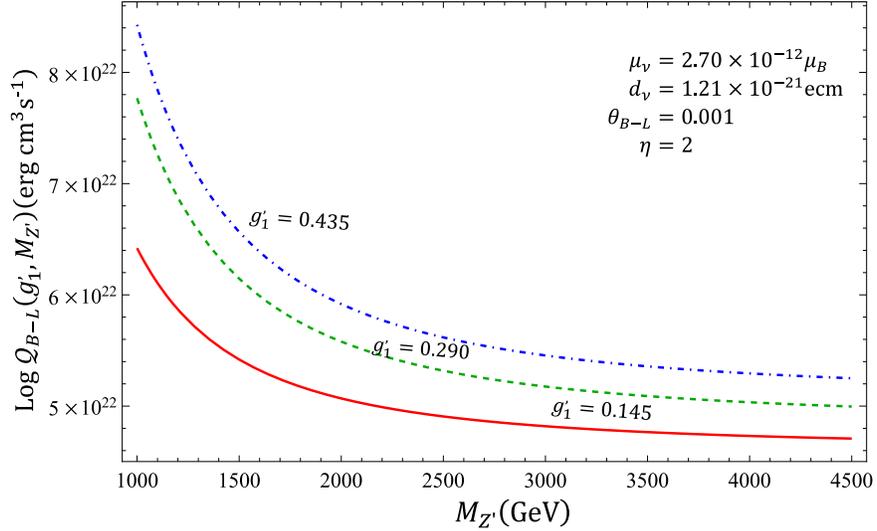}}}
	\caption{ \label{fig:Mz-g} The stellar energy loss rates
		as a function of the mass of the new gauge boson $M_{Z^{\prime}}$. The solid line is for $g^{\prime}_1=0.145$,
the dashes line is for $g^{\prime}_1=0.290$
		and the dot-dashes line is for $g^{\prime}_1=0.435$.}
\end{figure}

\begin{figure}[H]
	\centerline{\scalebox{.85}{\includegraphics{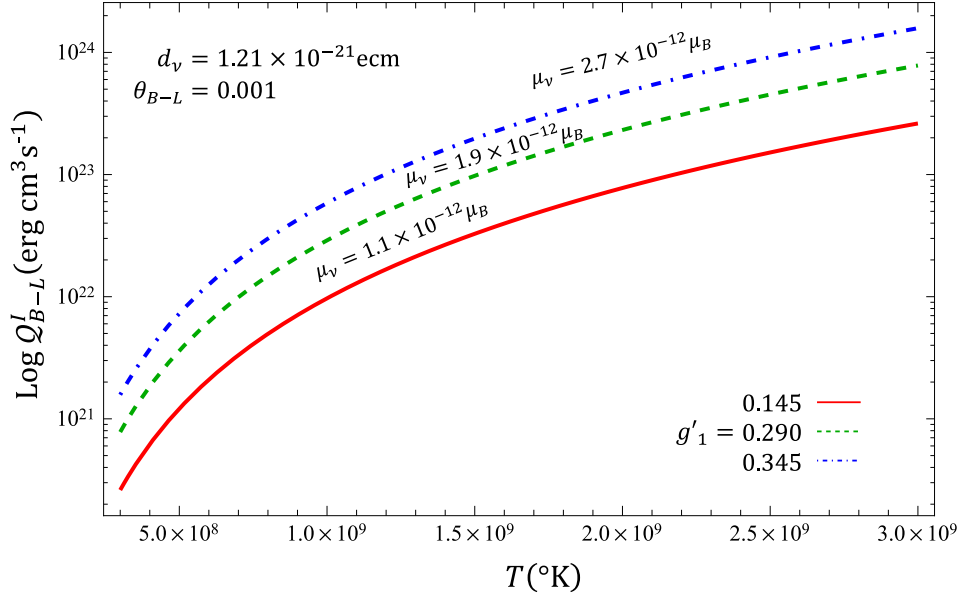}}}
	\caption{ \label{fig:region-1} The stellar energy loss rates
		as a function of the stellar temperature $T$. The solid line is for $g'_1=0.145$, the dashes line is for $g'_1=0.290$
and the dot-dashes line is for $g'_1=0.435$.}
\end{figure}

Fig. \ref{fig:Mz-g} shows the variation of the stellar energy loss rates as a function of the new gauge boson mass $M_{Z^{\prime}}$
from the $U(1)_{B-L}$ model with $g'_1=0.145, 0.290, 0.435$. A dependance on both $M_{Z^{\prime}}$ and $g'_1$ is observed for
${\cal Q}\left(M_{Z^{\prime}}, g'_1, \eta \right)$. This result comes directly from the relation $M_{Z^{\prime}}=2v^{\prime}g^{\prime}_1$ \cite{khalil2008,emam2007,basso2009}. In Fig. \ref{fig:region-1}, we show the stellar energy loss rates for the annihilation process
$e^{+}e^{-}\rightarrow (\gamma,Z,Z^{\prime}) \to \nu \bar \nu$ with contribution weak and electromagnetic. The high temperature
enhancement of the electromagnetic annihilation channel discussed in section II is evident. For an AMM of $10^{-12} \mu_B$ and
an EDM of $10^{-21} ecm$ the electromagnetic annihilation rates dominates over the rates for an AMM of $10^{-12} \mu_B$ and
an EDM of $10^{-21} ecm$, respectively. Clearly magnetic and electric dipole moments of order minor that of $10^{-12} \mu_B$
and $10^{-21} ecm$ have less influence on the annihilation channel.

A simple comparison of the SM and the $U(1)_{B-L}$ model for pair production $e^+e^-\to(\gamma, Z, Z^{\prime})\to\nu\bar\nu$
is obtained using the full expressions of $\mathcal{Q}_{B-L}\left(\mu_{\nu}, d_{\nu},\theta_{B-L},g^{\prime}_{1},M_{Z^{\prime}} \right)$
given by Eq. (18) and the corresponding of the SM which is obtained in the decoupling limit when $ \theta_{B-L}=0 $, $g^{\prime}_1=0$,
$M_{Z^{\prime}}\to \infty $ and $ \mu_{\nu}=d_{\nu}=0$.

\begin{figure}[H]
	\centerline{\scalebox{.7}{\includegraphics{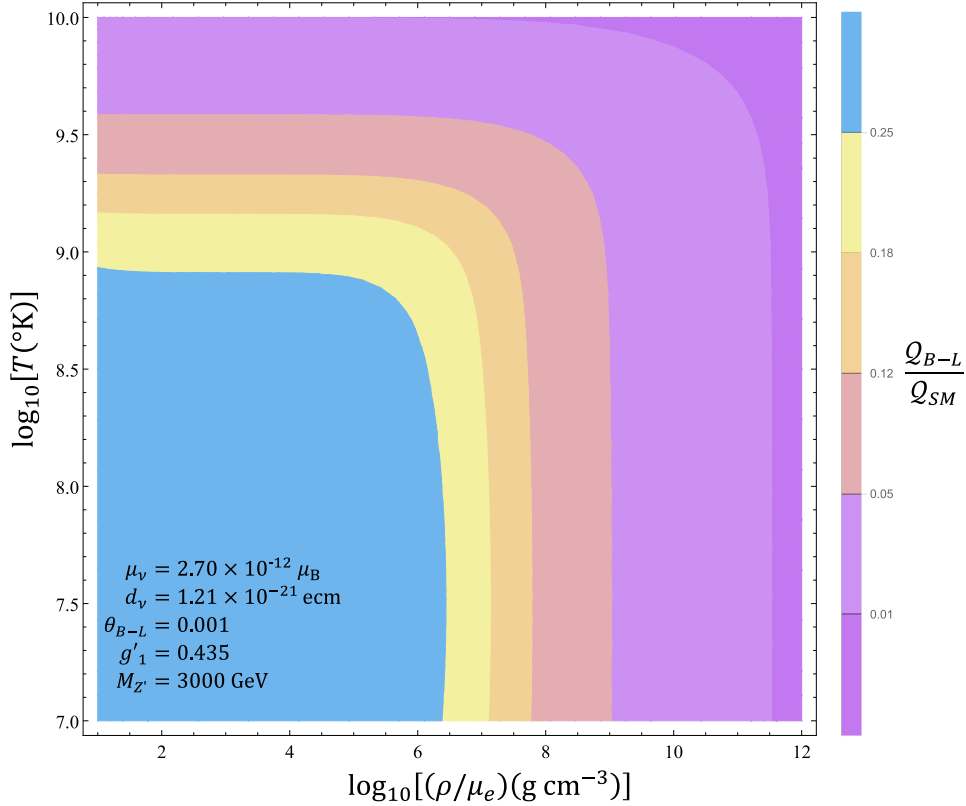}}}
	\caption{ \label{fig:T-rho} Contour plot of the ratio
    $\frac{\mathcal{Q}_{B-L}(\mu_{\nu}, d_{\nu}, \theta_{B-L}, g^{\prime}_{1}, M_{Z^{\prime}} )}{\mathcal{Q}_{SM} }$
     as a function of temperature $(T)$ and matter density $(\rho)$. }
\end{figure}

In this regard, Fig. \ref{fig:T-rho} shows the contour plot of the ratio $\frac{\mathcal{Q}_{B-L}(\mu_{\nu}, d_{\nu},
\theta_{B-L}, g^{\prime}_{1}, M_{Z^{\prime}} )}{\mathcal{Q}_{SM} }$ in the $(T, \rho/\mu_e)$ parameter space, where the input
parameters are $g^{\prime}_1=0.435$, $M_{Z^{\prime}}=3000$ GeV, $\theta_{B-L}=10^{-3}$, $\mu_{\nu}=2.70\times 10^{-12}\mu_B$
and $d_{\nu}=1.21\times 10^{-21} ecm$, respectively. This figure show the contours of the ratio $\mathcal{Q}_{B-L}\left(\mu_{\nu}, d_{\nu},\theta_{B-L},g^{\prime}_{1},M_{Z^{\prime}}\right)/\mathcal{Q}_{SM}$ of nonstandard over standard energy loss via neutrino
pair production as a function of temperature $(T)$ and matter density $(\rho)$ in the region of interest defined by $10^1 \leq
(\rho/\mu_e)(gcm^{-3})\leq 10^{12}$ and $10^7 \leq T(^0K) \leq 10^{10} $. The results show that indeed the $U(1)_{B-L}$ model
loss does not exceeds $25\%$ of the SM one.

\begin{figure}[H]
	\centerline{\scalebox{.7}{\includegraphics{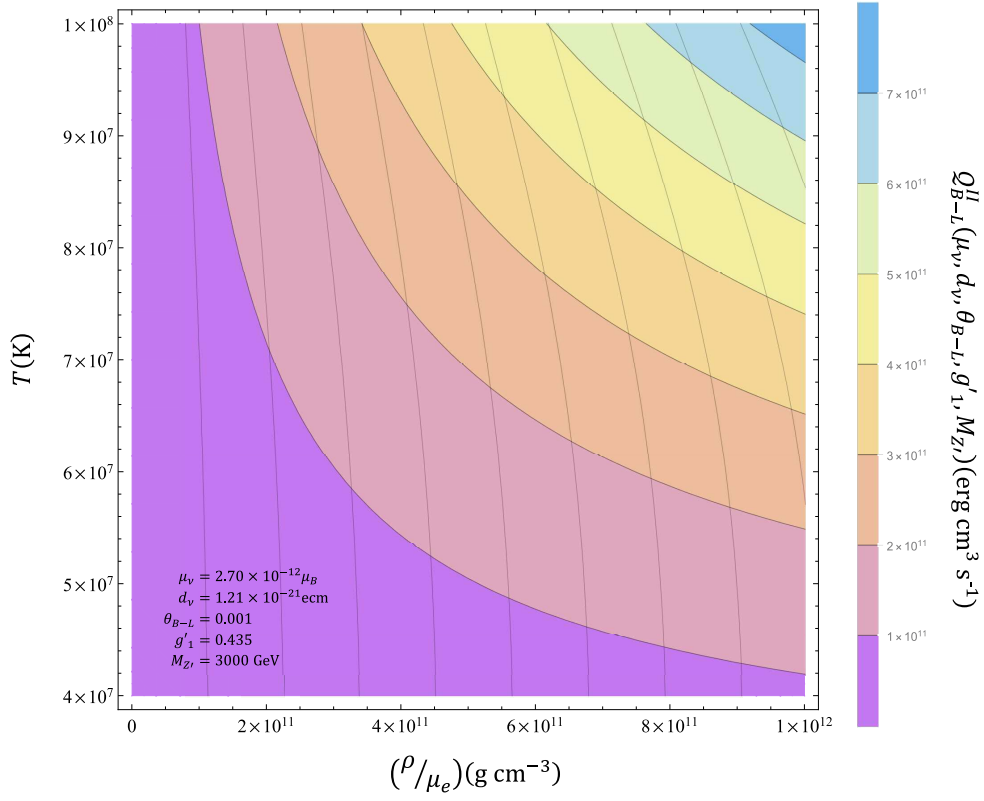}}}
	\caption{ \label{density} Contours in the plane $(T, \rho/\mu_e)$
for the Region II. The AMM of the neutrino is fixed at $\mu_\nu=2.70\times10^{-12} \mu_B$.}
\end{figure}

Starting from the expression for $ \mathcal{Q}^{B-L}_{II} $ given by Eq. \eqref{R2_1}, we shows the contours plot in the
$(T, \rho/\mu_e)$ plane for different $\mathcal{Q}_{II}^{B-L}\left(\mu_{\nu}, d_{\nu},\theta_{B-L},g^{\prime}_{1},M_{Z^{\prime}}
\right)$, as shown in the color code for $\mathcal{Q}_{II}^{B-L}\left(\mu_{\nu}, d_{\nu},\theta_{B-L},g^{\prime}_{1},M_{Z^{\prime}}
\right)$, and the AMM of the neutrino is fixed at $\mu_\nu=2.70\times10^{-12} \mu_B$. Our results are show in Fig. \ref{density}.

\begin{figure}[H]
	\centerline{\scalebox{.9}{\includegraphics{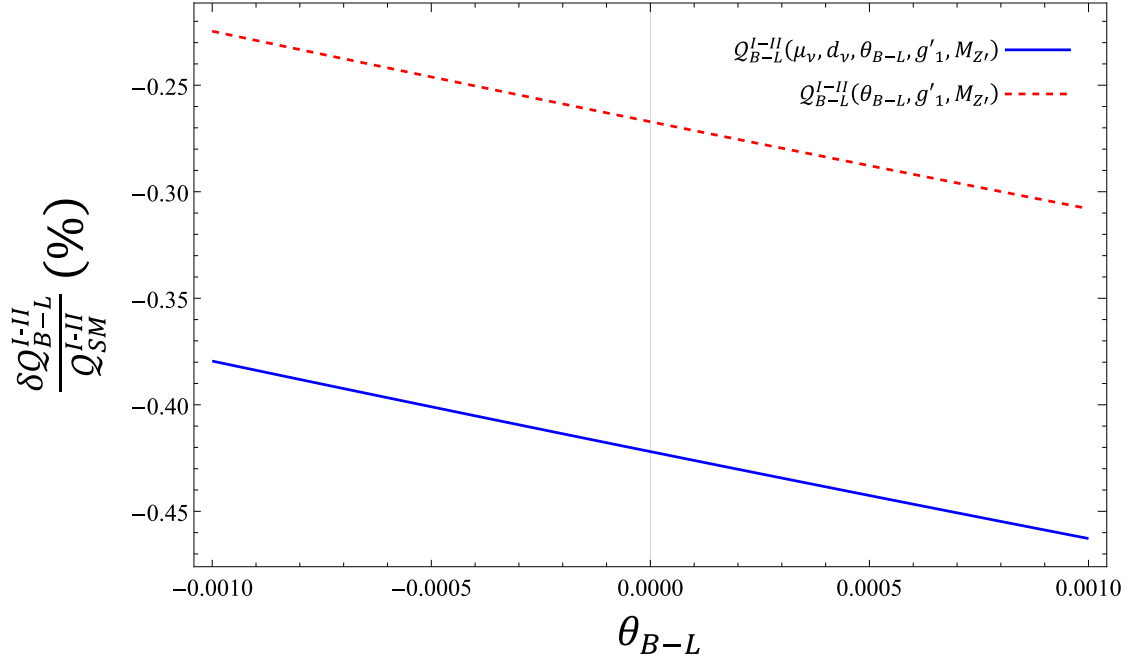}}}
	\caption{ \label{fig:region1-2} The relative correction
$\dfrac{\delta\mathcal{Q}_{B-L}^{I-II}  }{\mathcal{Q}_{SM}^{I-II} }$ as a function of the mixing angle $\theta_{B-L}$
with and without dependence of the AMM and the EDM $(\mu_{\nu}, d_{\nu})$ for Regions I and II. }
\end{figure}

\begin{figure}[H]
	\centerline{\scalebox{.9}{\includegraphics{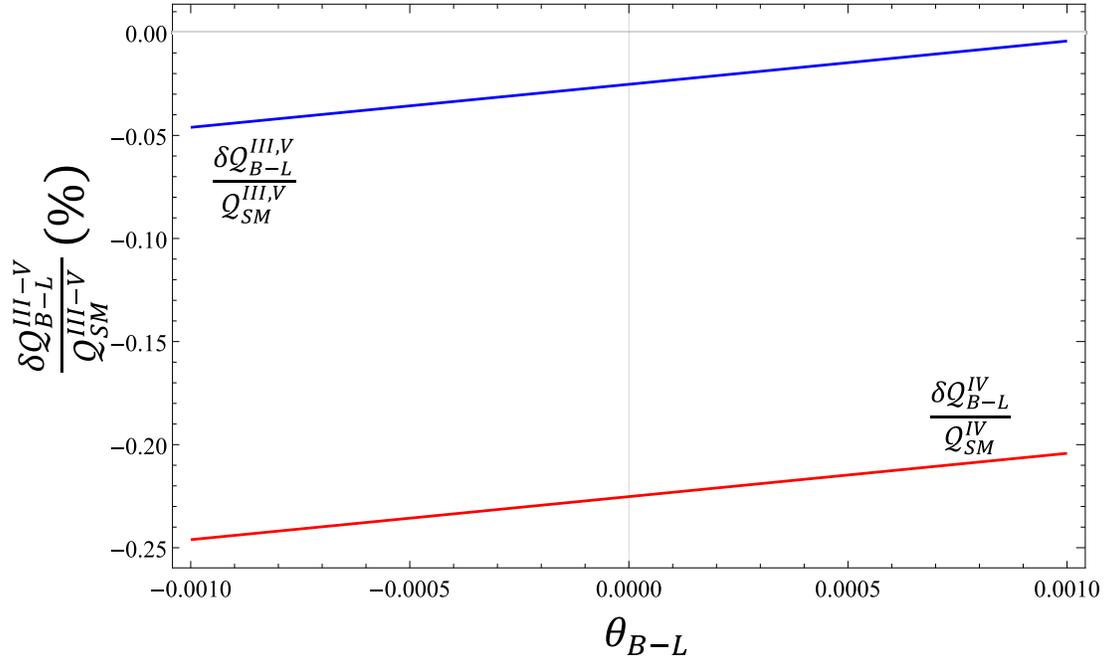}}}
	\caption{ \label{fig:region-3-5} The relative correction
		$\dfrac{\mathcal{Q}_{B-L}^{III-V}\left(\mu_{\nu}, d_{\nu},
			\theta_{B-L}, g^{\prime} _1,M_{Z^{\prime}},\eta  \right) - \mathcal{Q}_{SM}^{III-V}\left(\eta \right)}
        {\mathcal{Q}_{SM}^{III-V}\left(\eta \right) }$ as a function of the mixing angle $\theta_{B-L}.$}
\end{figure}

To visualize and quantify the effects of the dipole moments $\mu_{\nu}$ and $d_{\nu}$ of the neutrino, as well as of the parameters
$g^{\prime}_1$, $\theta_{B-L}$ and $M_{Z^{\prime}}$ of the $U(1)_{B-L}$ model on the stellar energy loss rates, we plot the relative
correction $\frac{\delta{\cal Q}^{B-L}_{\nu\bar\nu}}{ {\mathcal{Q}^{SM}_{\nu\bar\nu}}}$ for the regions I-II given by Eqs. \eqref{C_RI}
and \eqref{C_RII} in Fig. \ref{fig:region1-2}. From this figure we observed that the relative correction $\frac{\delta{\cal Q}^{B-L}_{I-II}}
{ {\mathcal{Q}^{SM}}}$ is of the order of $22\%-46\%$ to the interval of $-10^{-3}\leq \theta_{B-L} \leq 10^{-3}$. While for the regions III-V,
the relative correction $\frac{\delta{\cal Q}^{B-L}_{III-V}}{ {\mathcal{Q}^{SM}}}$ given by Eqs. \eqref{C_RIII}, \eqref{C_RIV} and \eqref{C_RV}
is of the order of $1\%-25\%$, as shown in Fig. \ref{fig:region-3-5}.

To show the consistency of our model with the minimally extended standard model, as well as with the role in cosmology and astrophysics
we estimated a sensitivity measure on the magnetic moment of the neutrino. Our obtained sensitivity measure is as follows. From Eqs. (21)
or (24) corresponding to regions I and II where there is dependence on the AMM $(\mu_{\nu})$ and considering the ratio:

\begin{eqnarray}
\frac{\mathcal{Q}^{B-L}_{I} }{\mathcal{Q}^{SM}_{I}}&=&\frac{M_Z^{4}\bigg[ \Big(g_{1}^{[B-L]}+g_{2}^{[B-L]}\Big)+\frac{\sin^4{2\theta_{W}}\left(\mu_\nu^2 + d_\nu^2 \right)}{16 \pi\alpha m_e^2}\bigg] }
{\Big( g_V^{SM}\Big)^2 },
\end{eqnarray}

\noindent it is possible to estimate sensitivity measure on the magnetic dipole moment of the neutrino which is competitive with
other bounds reported in different astrophysical, cosmological and experimental approaches, as the ones mentioned in the Introduction.
For instance, for the following values ​​of $ \theta_{B-L} = 10^{-3}$, $g^{\prime}_{1}=0.435 $ and $ M_{Z^{\prime}}=3000$ GeV we obtain
the following sensitivity measure for the AMM:

\begin{equation}
|\mu_\nu| \leq 2.020\times10^{-12}\mu_B.
\end{equation}

Our sensitivity measure for the AMM given by Eq. (36) is consistent with that obtained through the Supernova energy loss
studies lead to an upper limit of $\mu_\nu \lesssim (1.1-2.7)\times 10^{-12}\mu_B$  \cite{Kuznetsov-MM}. In addition of the limits
reported through the cooling rates of red giants $\mu_\nu \lesssim 3\times 10^{-12}\mu_B$ \cite{Raffelt-MM}, of the analysis of cooling
rates of white dwarfs $\mu_\nu \lesssim 10^{-11}\mu_B$  \cite{Blinnikov-MM}, etc..

We show that the stellar energy loss rates significantly depends on the effective magnetic moment of the neutrino, as well as on the
parameters of the $U(1)_{B-L}$ model, that is the mass of the new gauge boson $M_Z'$ , the mixing angle $\theta_{B-L}$ between $Z-Z'$
and the coupling constant $g'_1$. Furthermore, our analytical formulas for the stellar energy loss rates in the B-L model approach
are novel and are more general than the corresponding ones for the SM.

In the B-L model approach we estimated a sensitivity measure on the neutrino magnetic moment which is competitive with those obtained
from astrophysical and experimental approach. See Eq. (36) about our sensitivity measure on the AMM.

For all the aforementioned, ${\cal Q}^{B-L}_{\nu\bar \nu}$ increases if one introduces new interactions that changes the neutrino
annihilation cross-section. This is the case if the neutrino has a diagonal magnetic moment, because a magnetic moment would
increase $\nu-\bar\nu$ annihilation (creation) into (by) $e^{\pm}$, keeping the neutrinos in equilibrium below the canonical
(including only weak processes) neutrino decoupling temperature. In addition, as neutrino mass is sufficiently small, that is
$m_\nu \ll m_e$ and remains coupled to electrons while the electrons annihilate, the neutrino number density will be increased
because part of the electrons entropy will be shared with the neutrinos.

Until now, SN 1987A is the only supernova from which neutrinos have been detected. However, new generations of detectors of
neutrinos will increase the capability detection of the order of ten thousand neutrinos from Via Lactea supernovas up to one
more order of magnitude. As we just did, it will be possible to relate our results with data from observations or measurements
such as supernova energy loss, cooling rates of red giants, cooling rates of white dwarfs, Borexino Experiment, TEXONO
Collaboration, etc. through the neutrino magnetic moment.

\section{Conclusions}

We have developed and presented exact analytical formulas to assess the stellar energy loss rates involved in the process of emission
of neutrinos driven by the channel $e^+e^- \to (\gamma, Z, Z^{\prime}) \to \nu \bar\nu$. Thus, the stellar energy lost rates must be
assessed completely with \eqref{emissivity}. The validity of such an equation extends to the cases whether or not electrons are
degenerated or relativistic ones, a fact that is properly reflected in all the range of values considered for $\lambda$ and $\eta$.
A fiduciary result is that the stellar energy lost rates increase and are considerably dependent on the electromagnetic dipole moments
of the Dirac neutrinos and of the $U(1)_{B-L}$ parameters. The comparison $\mathcal{Q}_{B-L}\left(\mu_{\nu}, d_{\nu},\theta_{B-L},
g^{\prime}_{1},M_{Z^{\prime}} \right)/\mathcal{Q}_{SM}$ in the region of interest for $T$ and $\rho/\mu_e$ show that the $U(1)_{B-L}$
model loss does not exceeds $25\%$ of the SM one. Also, we find that the sensitivity estimated on magnetic dipole moment of the neutrino
is of the order of $|\mu_\nu| \leq 2.020\times10^{-12}\mu_B$, and it is competitive with other limits reported in the literature
\cite{Barbieri-MM,Kuznetsov-MM,Raffelt-MM,Blinnikov-MM,Borexino-MM,Texono-MM}. The relative correction of $\frac{\delta{\cal Q}^{B-L}_{\nu\bar\nu}}
{ \mathcal{Q}^{SM}_{\nu\bar\nu}}$, given by $\frac{\delta\mathcal{Q}^{B-L}_{I-II}}{ {\mathcal{Q}^{SM}}}$ is of $22\%-46\%$ for
the regions I-II. For regions III-V the relative correction is of the order of $1\%-25\%$. The comparison $\mathcal{Q}_{B-L}\left(\mu_{\nu}, d_{\nu},\theta_{B-L},g^{\prime}_{1},M_{Z^{\prime}} \right)/\mathcal{Q}_{SM}$ in the region of interest for $T$ and $\rho/\mu_e$ show that the
$U(1)_{B-L}$ model loss does not exceeds $25\%$ of the SM one.

With our results, the process of pair-annihilation $e^{+}e^{-}\rightarrow (\gamma, Z, Z^{\prime}) \to \nu \bar \nu$, open a number
of opportunities to further study the stellar energy loss rates combining both, the effects of the AMM and the EDM of the neutrino,
and the $U(1)_{B-L}$ parameters, with the inclusion of other potentially important channnels such as $\gamma + e^{\pm}\rightarrow
e^{\pm}\nu \bar \nu$  ($\nu$- photoproduction), $\gamma^* \rightarrow \nu \bar \nu$ (plasmon decay) and $e^{\pm} +Z \rightarrow
e^{\pm} +Z \nu \bar \nu $ (bremsstrahlung on nuclei). These processes are the dominant cause of the stellar energy loss rates in
different regions present within the density-temperature plane. These new calculations, could contribute to a better understanding
of the neutrino physics, and of new physics BSM \cite{Gutierrez2019}.

\vspace{1cm}

\begin{center}
	{\bf Acknowledgments}
\end{center}

A. G. R. and M. A. H. R. thank SNI and PROFEXCE (M\'exico).

\vspace{1cm}

\appendix

\section{Lagrangian of the $U(1)_{B-L}$ model}

For the Lagrangian of the $U(1)_{B-L}$ model, the terms for the interactions between
neutral gauge bosons $Z, Z^{\prime}$ and a pair of fermions of the SM can be written as
\cite{khalil2008,emam2007,bugarin2017}:

\begin{equation}
{\cal L}_{NC}=\frac{-ig}{\cos\theta_W}\sum_f\bar f\gamma^\mu\frac{1}{2}(g^f_V- g^f_A\gamma^5)f Z_\mu + \frac{-ig}{\cos\theta_W}\sum_f\bar f\gamma^\mu\frac{1}{2}(g^{\prime f}_V- g^{\prime f}_A\gamma^5)f Z^{\prime}_\mu.
\end{equation}

Thus, the expressions for the new couplings between the $Z, Z^{\prime}$ bosons and the SM fermions are presented in Table I. As usual,
the SM couplings are recovered in the limit when $\theta_{B-L}=0$ and $g^{\prime}_1=0$,

\begin{table}[!ht]
	\caption{New couplings of the $Z, Z^{\prime}$ bosons with the SM fermions.
		$g=e/\sin\theta_W$ and $\theta_{B-L}$ is the $Z-Z^{\prime}$ mixing angle.}
	\begin{center}
		\begin{tabular}{|c|c|}
			\hline\hline
			\hspace{2mm}Particle       & \hspace{5mm}                Couplings                       \\
			\hline\hline
			\hspace{2mm} $f\bar f Z$    & \hspace{5mm}      $g^f_V=T^f_3\cos\theta_{B-L}-2Q_f\sin^2\theta_W\cos\theta_{B-L}+\frac{2{g\prime}_1}{g}\cos\theta_W \sin\theta_{B-L},$ \hspace{2mm}  \\
			
			&      \hspace{5mm} $g^f_A=T^f_3\cos\theta_{B-L}$\hspace{2mm}                                                                                 \\
			\hline
			
			\hspace{2mm} $f\bar f Z^\prime$   & \hspace{5mm}      ${g^\prime}^{f}_V=-T^f_3\sin\theta_{B-L}-2Q_f \sin^2\theta_W \sin\theta_{B-L}+\frac{2{g^\prime}_1}{g}\cos\theta_W \cos\theta_{B-L},$\hspace{2mm} \\
			
			& \hspace{5mm}      ${g\prime}^{f}_A=-T^f_3\sin\theta_{B-L}$\hspace{2mm}\\
			\hline\hline
		\end{tabular}
	\end{center}\label{tabla:couplings}
\end{table}

\section{Couplings Constants}

In Eq. (18) we have redefined the coupling constants of the $U\left(1\right)_{B-L}$ model as:

\begin{align}
g_{1}^{[B-L]}&=\Bigg[\dfrac{1}{M_Z^4}\Big( \left(g_V^{e}\right)^2 +\left(g_A^{e}\right)^2 \Big)\Big( \left(g_V^{\nu}\right)^2 +\left(g_A^{\nu}\right)^2 \Big)+\dfrac{1}{M_{Z^\prime}^4}\Big( \left(g_V^{\prime e}\right)^2 +\left(g_A^{\prime e}\right)^2 \Big)\nonumber \\
&\qquad\times\Big( \left(g_V^{\prime\nu}\right)^2 +\left(g_A^{\prime\nu}\right)^2 \Big) +\dfrac{2}{M_Z^2M_{Z^\prime}^2}\big(g_V^{e}g_V^{\prime e}+g_A^{e}g_A^{\prime e}\big)\big( g_V^{\nu}g_V^{\prime \nu}+g_A^{\nu}g_A^{\prime \nu}\big)\Bigg],
\end{align}

\begin{align}
g_{2}^{[B-L]}&=\Bigg[\dfrac{1}{M_Z^4}\Big( \left(g_V^{e}\right)^2 -\left(g_A^{e}\right)^2 \Big)\Big( \left(g_V^{\nu}\right)^2 +\left(g_A^{\nu}\right)^2 \Big)+\dfrac{1}{M_{Z^\prime}^4}\Big( \left(g_V^{\prime e}\right)^2 -\left(g_A^{\prime e}\right)^2 \Big)\nonumber \\
&\qquad\times\Big( \left(g_V^{\prime\nu}\right)^2 +\left(g_A^{\prime\nu}\right)^2 \Big) +\dfrac{2}{M_Z^2M_{Z^\prime}^2}\big(g_V^{e}g_V^{\prime e}-g_A^{e}g_A^{\prime e}\big)\big( g_V^{\nu}g_V^{\prime \nu}+g_A^{\nu}g_A^{\prime \nu}\big)\Bigg].
\end{align}



\vspace{2cm}

\end{document}